\documentclass[letter]{aa}  
\usepackage{graphicx}
\usepackage{natbib}
\usepackage{txfonts}
\newcommand{\Msun}{\ensuremath{\,{\rm M}_{\odot}}}
\newcommand{\msun}{\ensuremath{\,{\rm M}_{\odot}}}

\def\araa{ARA\&A}             % Annual Review of Astron and Astrophys
\def\apj{ApJ}                 % Astrophysical Journal
\def\apjl{ApJ~Lett.}          % Astrophysical Journal, Letters
\def\aap{A\&A}                % Astronomy and Astrophysics
\def\mnras{MNRAS}             % Monthly Notices of the RAS
\def\nat{Nature}              % Nature
\begin{document}

   \title{Fluorine in carbon-enhanced metal-poor stars: a binary scenario}

   \author{M. Lugaro
          \inst{1,2}\fnmsep\thanks{The first two authors have contributed 
equally to this paper.}
          \and
          S. E. de Mink
          \inst{1,\star}
          \and
          R. G. Izzard
          \inst{1}
          \and
	  S. W. Campbell
          \inst{3,2}
          \and
	  A. I. Karakas
          \inst{4}
          \and
          S. Cristallo
          \inst{5}
          \and
          O. R. Pols
          \inst{1}
          \and
          J. C. Lattanzio
          \inst{2}
          \and
          O. Straniero
          \inst{5}
          \and
          R. Gallino
          \inst{6,2}
          \and
          T. C. Beers
          \inst{7}
          }

 %latexps   \offprints{M. Lugaro, m.a.lugaro@uu.nl}

   \institute{Sterrekundig Instituut, Universiteit Utrecht, P.O. Box 80000,
         3508 TA Utrecht, The Netherlands\\
         \email{m.a.lugaro@uu.nl,s.e.demink@uu.nl,r.g.izzard@uu.nl,o.r.pols@uu.nl}
         \and
         Centre for Stellar and Planetary Astrophysics, School of Mathematical Sciences,
         Monash University, Victoria 3800, Australia\\
         \email{John.Lattanzio@sci.monash.edu.au}
         \and
         Academia Sinica Institute of Astronomy \& Astrophysics, Taipei, Taiwan\\
         \email{simcam@asiaa.sinica.edu.tw}
         \and
         Research School of Astronomy and Astrophysics, Mt. Stromlo Observatory, 
         Cotter Rd., Weston, ACT 2611, Australia\\   
         \email{akarakas@mso.anu.edu.au}
         \and 
         INAF, Osservatorio Astronomico di Collurania, 64100 Teramo, Italy\\
         \email{cristallo@oa-teramo.inaf.it}
         \email{straniero@oa-teramo.inaf.it}
         \and
         Dipartimento di Fisica Generale, Universit\'a di Torino, Torino, Italy\\ 
         \email{gallino@ph.unito.it}
         \and
         Department of Physics and Astronomy, Center for the Study of Cosmic 
         Evolution, and Joint Institute for Nuclear Astrophysics, Michigan State 
         University, East Lansing, MI, USA\\
         \email{beers@pa.msu.edu}
}

   \date{Received -- , 2007; accepted -- --, 2007}

   \authorrunning{Lugaro and De Mink et al.}
   \titlerunning{Fluorine in CEMP stars}

   \abstract {} {A super-solar fluorine abundance was 
     observed in the carbon-enhanced metal-poor (CEMP) star
     HE~1305+0132 ([F/Fe] = $+$2.90, [Fe/H] = $-$2.5). We
propose that this observation can be explained using
     a binary model that involve mass transfer from an asymptotic
     giant branch (AGB) star companion 
     and, based on this model, we predict F abundances in
     CEMP stars in general. We discuss wether F can be used to
     discriminate between the formation histories of most CEMP
     stars: via binary mass transfer or from the ejecta of
     fast-rotating massive stars.}
   {We compute AGB yields using different stellar evolution and
     nucleosynthesis codes to evaluate stellar model uncertainties. 
     We use a simple dilution model to 
     determine the factor by which the AGB yields should be diluted to
     match the abundances observed in HE~1305+0132.  We further employ
     a binary population synthesis tool to estimate the probability of
     F-rich CEMP stars.}
   {The abundances observed in HE~1305+0132 can be explained if this star
accreted 3-11\% of the mass lost by its former AGB companion. 
The primary AGB star should have dredged-up at least
     0.2\Msun~of material from its He-rich region into the convective 
     envelope via
     third dredge-up, which corresponds to AGB models of $Z \simeq 0.0001$
     and mass $\simeq$ 2\Msun. Many AGB model uncertainties, such as the
     treatment of convective borders and mass loss, require further
     investigation.  We find that in the binary scenario most CEMP
     stars should also be FEMP stars, that is, have [F/Fe] $>$ +1, while 
     fast-rotating massive
     stars do not appear to produce fluorine. We conclude that fluorine is a 
     signature of low-mass AGB pollution in CEMP stars, together 
     with elements associated with the $slow$ neutron-capture process.
}
          {}


   \maketitle

% ================================================================= 
% INTRO 
% ================================================================= 
\section{Introduction}\label{sect:intro}

Carbon-enhanced metal-poor (CEMP) stars are chemically peculiar
objects, which represent 10-20\% of all halo stars
\citep{beers05,cohen05c,lucatello06}. Most of CEMP stars exhibit radial velocity
variations, which imply the presence of a binary companion
\citep{lucatello05b}. A significant fraction of CEMP 
stars \citep[$\sim $70-80\%, according
to][CEMP-s]{aoki07} also exhibit enhancements in heavy elements such as Ba
and Pb, which are produced by $slow$ neutron captures ($s$ process) in
asymptotic giant branch (AGB) stars
\citep[e.g.,][]{gallino98}. One scenario to explain the 
abundance patterns in CEMP stars is therefore mass transfer from a
former AGB companion in which the carbon and heavy neutron-capture
elements were produced \citep[e.g.,][]{ivans05,thompson08}. However, a certain 
fraction of CEMP stars, which have typically [Fe/H]\footnote{[X/Y] = log(X/X$_{\odot}$) $-$ 
log(Y/Y$_{\odot}$).}$<-$2.7, 
exhibit low or no neutron-capture element
abundances (CEMP-no). These stars might have formed instead from material ejected
by rapidly rotating massive stars \citep{meynet06} or faint Type II 
supernovae \citep{umeda05}. At extremely low metallicities, [Fe/H] $< -$4,
giant CEMP stars could have enriched themselves in carbon via a ``dual
core flash'' - where mixing of protons during the core helium flash 
induces a hydrogen flash - while in the early phases of AGB 
stars of masses $\leq$ 1.5 \msun\ and [Fe/H] $\leq -$2.3, a ``dual 
shell flash'' may occur, where protons 
are ingested into the convective pulse 
\citep{fujimoto90,hollowell90,fujimoto00,picardi04,cristallo07}. 

\citet{schuler07} derived a super-solar fluorine abundance
of A(F)\footnote{A(Element) = log N(Element)+12.} = $+4.96 \pm 0.21$
for the halo star 
HE~1305+0132, which corresponds to 
[F/Fe] = $+$2.9. This is the
most Fe-deficient star, [Fe/H] = $-2.5 \pm 0.5$, for which the fluorine
abundance has been measured to date. HE~1305+0132 also exhibits
overabundances of C and N ([C/Fe] = $+2.68 \pm 0.51$;[N/Fe] = $+1.6 \pm 0.46$)
and an O abundance typical of halo stars ([O/Fe] = $+0.50 \pm 0.22$). 
Lines of Ba and Sr are observed
in its spectra \citep{goswami05}, which place HE~1305+0132 in the 
group of CEMP-s stars.

Fluorine abundances were first determined in AGB stars by \citet{jorissen92}. 
Enhancements of up to 30 times the
solar value were reported, demonstrating that these stars produce
fluorine. Observations of post-AGB stars and planetary nebulae,
the progeny of AGB stars, confirm that these objects are also enriched
in fluorine \citep{werner05,zhang05}. 
Type II supernovae \citep{woosley88} and Wolf-Rayet stars during helium burning
\citep{meynet00} have been theoretically identified as F production sites,
but they are not observationally confirmed. In contrast to that observed 
in CEMP stars and in particular HE~1305+0132, Type II 
supernovae typically produce more 
O than C, a part from a narrow range of initial mass around 80\msun\ 
\citep{woosley02} or in 
the case of faint supernovae at Z = 0 \citep{umeda05}, while  
models of Wolf-Rayet stars show that
fluorine production in these stars scales with stellar metallicity
and decreases when rapid rotation is included \citep{palacios05}.

\begin{figure}
  \centering 
  \includegraphics[angle=-90, width=0.5\textwidth]{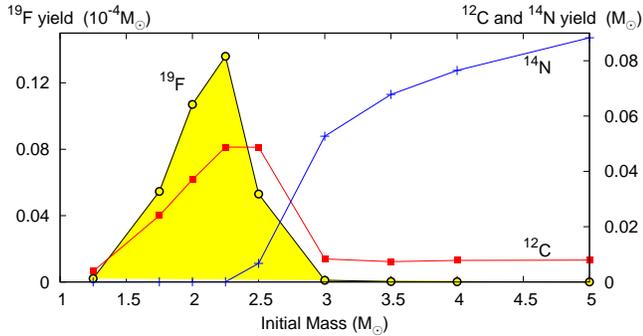}%
  \caption{C, N, and F yields from AGB models of different masses and 
[Fe/H]$ = -2.3$. Note that the F yields
    are plotted on a different scale.}
  \label{fig:yields}
\end{figure}

The aims of this paper are to discuss the yields of C, N, and F from AGB
stars at the metallicities relevant to CEMP stars (Sect. 2); to determine if
the abundances measured for HE~1305+0132 can be explained using the AGB
binary scenario (Sect. 3); and to evaluate the consequences of fluorine
production in AGB stars on the CEMP stellar population (Sect. 4).  We
evaluate AGB modelling uncertainties related to different
physics prescriptions and nuclear reaction rates in Sect. 5 and 
present our conclusions in Sect. 6.

\section {C, N, and F from low metallicity AGB stars}

Fluorine can be produced in AGB stars via the
$^{18}$O(p,$\alpha$)$^{15}$N($\alpha,\gamma)^{19}$F reaction chain
during thermal pulses (TPs) associated with the periodic activation of
the He-burning shell. While $^{18}$O is provided by the
$^{14}$N($\alpha,\gamma)^{18}$F($\beta^+\nu)^{18}$O chain, protons are
produced via $^{14}$N($n,p$)$^{14}$C reactions, and neutrons originate 
from $^{13}$C($\alpha,n)^{16}$O reactions. The $^{13}$C and $^{14}$N
nuclei are mixed down into the convective He shell from the ashes of
the H-burning shell. After each TP quenches, third dredge-up (TDU) may
occur, which trasports $^{19}$F to the convective envelope, together with
$^{12}$C produced by partial He burning in the TPs. 

Yields for C, N, and F at [Fe/H] = $-$2.3 from the models of  
\citet{lugaro04} and \citet{karakas07b} are shown in Fig.~\ref{fig:yields}. The 
reader is 
referred to these papers for details of the computational methods. The 
profiles of the C and the F yields as a function of the initial stellar 
mass closely follow each other. This is because the TDU  
carries primary $^{12}$C to the stellar envelope. This is 
converted into primary $^{13}$C and $^{14}$N in the H-burning ashes, 
whose abundances drive the synthesis of F.
Hence, fluorine production in AGB stars of low metallicity is of 
a primary nature \citep{lugaro04}. 
When the stellar mass is higher than $\simeq$ 
3 \msun, proton captures at the base of the convective envelope (hot bottom
burning, HBB) lead to the conversion of C into N and to the
destruction of F. Hence, we expect that the high F abundance 
([F/Fe] $\approx +$4) predicted by \citet{meynet06} in the envelope of 
a low-metallicity rotating 7 \msun\ model at the beginning of the AGB phase 
will be completely destroyed by HBB during the subsequent AGB phase.

Since the $^{12}$C/$^{13}$C ratio in HE~1305+0132 is observed close to 
its equilibrium value \citep{goswami05} and N is mildly enhanced, proton captures 
may have occurred inside this star, or in the AGB 
companion, due to non-convective-mixing processes that occurr below the inner 
boundary of the convective envelope. These can be due, for example, to 
efficient activation of thermohaline mixing \citep{charbonnel07}. 
An alternative explanation may be the occurrence of the dual shell flash in the AGB star.
The high observed F abundance indicates that F cannot have been significantly destroyed by 
proton captures 
via the $^{19}$F($p,\alpha$)$^{16}$O reaction, thus constraining the temperature 
reached by any extra mixing process to less than $\simeq$ 20 million degrees, 
within reaction rate uncertainties \citep[see, e.g, Fig. 2 and 3 of][]{arnould99}.

%\vspace{-0.4cm}

\section{Estimate of binary transfer parameters}

\begin{figure}
  \centering
  \includegraphics[angle=90,width=0.5\textwidth]{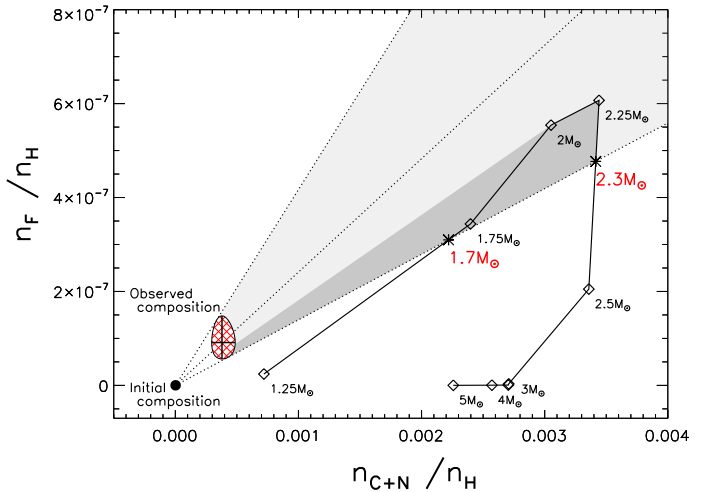}
  \caption{Abundances by number of F and C+N with respect to H as
    observed in HE~1305+0132 with 1 $\sigma$ errors, taken from
    \citet[hatched ellipsoid]{schuler07}, and computed 
    in the average material lost via the winds by AGB stars  
    at [Fe/H] = $-2.3$, with labels indicating the initial
    masses. The initial composition is scaled from solar. 
    The gray area is the region of the plot where the 
    F and C+N abundances from the AGB companion should lie to ensure that 
    the observed
    abundances are reproduced after dilution. The darker gray 
area represents the region
    covered by our models with masses in the range indicated by the asterisks.  
  }
  \label{fig:dilution} \end{figure}

In Fig.~\ref{fig:dilution} we show the abundances of F and C+N,
\footnote{We prefer to use C+N rather than C because our models do 
not take into account the possibility that some C is converted into N.} 
relative to H, observed in the star and compare these values 
to the abundance ratios in the 
material lost by AGB stars according to the models presented in 
Fig.~\ref{fig:yields}.
To reproduce the observed abundances we employ two free parameters: (1)
the initial mass of the polluting AGB star, which determines the chemical
composition of the accreted material, and (2) the amount by which the
accreted material is diluted into the envelope of the polluted
star. In Fig.~\ref{fig:dilution} the composition that is produced by the 
mixing of material before accretion with 
material from the AGB star lies on a straight line connecting
the two components; the amount of dilution 
is represented by the position of the point on the mixing line.
To explain the measured abundances, we require an AGB initial mass 
of between $1.7$ and $2.3$\Msun, and dilution of the accreted material 
by a factor of between six and nine.

If we assume that the mass of HE~1305+0132 is 0.8\Msun, which is the
typical mass of halo 
stars, and calculate the evolutionary track of a star of this
mass and metallicity Z = 10$^{-4}$ \citep[using the stellar evolution code STARS, see, 
e.g.,][]{pols95}, we find that the observed effective temperature
$T = 4.46 \pm 0.10$~kK \citep{schuler07} indicates that the star is a
giant with a convective envelope.  When the envelope reaches its
maximum depth, the outermost 60\% of the mass of the star is 
convective.\footnote{The mixed
fraction might be as much as 90\% if thermohaline mixing is
effective \citep{stancliffe07a}, even though this may be debated in
particular regarding the possible counter-effect of gravitational
settling \citep[see, e.g.,][]{thompson08,aoki08}.}
With these assumptions, we find that the star should have accreted
0.05-0.12\Msun~from a former AGB companion. Given that the total mass
lost during AGB evolution is in the range 1.0-1.5\Msun, this corresponds to
the accretion of 3-11\% of the mass lost by the AGB star. 

We implicitly assume that the composition of the accreted
material can be represented well by the average composition of the
total material ejected by the AGB star. In reality, however, the binary orbit
is altered during mass transfer, so the composition of the accreted
material varies because the surface AGB composition varies. Moreover, the 
evolution of the AGB star itself may be altered by the presence of a binary
companion. Since there are no models presently available that 
describe both binary and detailed AGB evolution simultaneously, we 
have defaulted to using the AGB average composition.

In principle, it is possible to constrain the initial orbital period
of the binary system. 
The system must have been sufficiently wide to allow the primary star to 
evolve into an AGB star before filling its Roche lobe; this 
requirement sets a lower limit to the initial orbital period. 
On the other
hand, the efficiency of wind accretion decreases with the distance
between the two stars, which implies that the required amount of 
accreted material
may set an upper limit on the initial orbital period. Using the binary
evolution code described in \citet{izzard06}, we find that the range of
initial orbital periods that corresponds to an accretion efficiency $>3\%$ in
a system with initial masses 0.8 and 2\Msun~is between 7 and 27 years.

\section{Fluorine abundances in CEMP stars from binary systems}

With the binary evolution code described in \citet{izzard06}, we
simulated a population of binaries assuming the \citet{kroupa93}
initial mass function, a uniform distribution of initial mass
ratios, and the \citet{duquennoy91} initial period distribution.

We find that, of all turn-off and giant CEMP stars formed by mass
transfer from an AGB companion, 81\% are expected to show an enhanced
fluorine abundance of [F/Fe] $>$ +1, 12\% have [F/Fe] $>$
+2, and 0.005\% have [F/Fe] $>$+3. This can be understood qualitatively
from the results presented in Fig.~\ref{fig:yields} because 
AGB stars that produce carbon also produce
fluorine. To date, fluorine enhancements have not been reported for other
CEMP stars because high-resolution spectra in the 2.3-2.4 $\mu$m band
are required, which are not yet
available. We predict that fluorine should be found in most CEMP stars
that formed via the AGB binary scenario, thus representing a tracer of low-mass
AGB pollution in addition to $s$-process element enhancements.

The fluorine abundance in the particular case of HE~1305+0132 appears to
be exceptionally high, since we expect 
only 0.04\% of turn-off and giant CEMP stars to have A($^{19}$F)$ \geq 
+4.75$, when considering the observational error bar.
However, we emphasize that many uncertainties play a role in this estimate,
both in the assumed distribution functions (which are reasonable for
stars in the solar neighbourhood, but not necessarily for halo stars),
and in the assumed wind accretion efficiency. The latter is based on
the \citet{bondi44} prescription, whereas hydrodynamical simulations
\citep{theuns96,nagae04} predict typically lower accretion efficiencies. 
These uncertainties deserve more attention in a follow-up
study (Izzard et al., in preparation). Moreover, 
the theoretical understanding of
F production in
AGB stars is itself still affected by many uncertainties, as discussed below.

\begin{table}[b]
\begin{center} 
  \caption{Number of TPs with TDU, total
    TDU mass, and C, N, and F yields (all in \Msun) for different AGB
    models of 2\Msun~at [Fe/H]$ = -2.3$. \label{tab:uncertain_yields} }
\begin{tabular}{llccccc} 
\hline 
& Model & TPs & M$_{\rm TDU}$ & C & N & F \\ 
\hline 
1&Karakas$^a$ & 25 & 0.22 & 3.7E-02 & 7.2E-05 & 1.1E-05 \\ 
2&pocket$^b$ & " & " & " & " & 1.7E-05 \\ 
3&$^{18}$F($\alpha,p$) UL$^b$ & " & " & " & " & 2.8E-05 \\ 
4&pocket + UL$^b$ & " & " & " & " & 3.3E-05 \\ 
%5&Campbell$^c$ & 125 & 0.03 & 4.5E-04 & 6.3E-03 & 7.1E-09 \\ 
5&Cristallo$^c$ & 14 & 0.09 & 1.7E-02 & 3.4E-05 & 2.4E-06 \\ 
6&Cristallo$^d$ & 49 & 0.19 & 3.1E-2 & 1.2E-4 & 1.1E-05 \\ 
\hline 
\end{tabular} 
\end{center} 
$^a$\citet{lugaro04} and \citet{karakas07b}. 
$^b$\citet{karakas08}. 
%$^c$This work.
$^c$\citet{cristallo07}. $^d$\citet{cristalloThesis}.
 
\end{table}

\section{Uncertainties in AGB models}

To evaluate AGB model uncertainties we discuss a set of models of
2\Msun~and [Fe/H] = $-2.3$, computed using different physics and nuclear
reaction rate assumptions (Table~\ref{tab:uncertain_yields}). The
first four models in Table~\ref{tab:uncertain_yields} are 
computed using the codes described
in \citet{karakas07b}. Model 1 is that used in
the previous section for comparison with HE~1305+0132. Model 2 includes a region in 
which protons from the envelope are mixed
down into the top layer of the He- and C-rich intershell (the region
between the H and He shells) at the end of each TDU. Proton captures
on $^{12}$C generate a ``pocket'' rich in $^{13}$C, the main 
neutron source for the $s$-process in these stars, and $^{14}$N
\citep{herwig05}. Extra $^{15}$N is
produced in the $^{13}$C-$^{14}$N pocket, which is then converted to
$^{19}$F in the following convective TP \citep[see
also][]{goriely00}. The introduction of a $^{13}$C-$^{14}$N pocket of
0.002\Msun~increases the $^{19}$F yield by 60\%, while the C and N
yields are unaffected. In Model 3 of Table~\ref{tab:uncertain_yields}
we consider a model computed using the upper limit
(UL) of the $^{18}$F($\alpha,p$)$^{21}$Ne rate. This increases the
$^{19}$F yield by a factor of $\simeq$2.5 \citep[see details and
discussion in][]{karakas08}. When both a $^{13}$C-$^{14}$N pocket and
the upper limit of the $^{18}$F($\alpha,p$)$^{21}$Ne reaction rate is
used (Model 4 of Table~\ref{tab:uncertain_yields}), the $^{19}$F yield
increases by a factor of three. The other
main nuclear uncertainties originate from the 
$^{14}$C($\alpha,\gamma$)$^{18}$O
and the $^{19}$F($\alpha,p$)$^{22}$Ne reaction rates \citep[see discussion 
in][]{lugaro04}. For 
the latter, the latest evaluation by \citet{ugalde08} needs 
to be tested in AGB models.

All models presented do not take account of the effects
induced by carbon enhancement on the opacities of the cool external
layers of AGB stars.  When C/O $>$ 1, C-bearing molecules, most 
notably C$_2$
and CN, increase the opacity of the external layers, causing the
envelope to expand and the star to become larger and
cooler \citep{marigo02}.  Models of AGB stars of low mass and
metallicity with C- and N-enhanced low-temperature opacities have been 
calculated using the Frascati Raphson Newton evolutionary code (FRANEC)
code \citep{straniero06,cristallo07}. In these models the
mass-loss rate strongly increases with respect to models in which
opacities are always calculated using the initial $Z = 10^{-4}$
solar-scaled composition. The resulting yields (Model 5 of
Table~\ref{tab:uncertain_yields}) are $\approx$ 5 times smaller than
in the Karakas models. For comparison, the results obtained with the
same code, but using opacities calculated for the initial solar-scaled
composition \citep{cristalloThesis} are also reported in
Table~\ref{tab:uncertain_yields} (Model 6), and are in good agreement
with the Karakas model, in spite of the different choices of
mass-loss rate and treatment of the convective borders
\citep[see][for details]{straniero06}.

It is evident that further work is required to address the uncertainties
in the AGB fluorine yields at low metallicities. In particular, the
inclusion of low-temperature carbon-enhanced opacities in the Karakas
models (Karakas, Wood \& Campbell, in preparation) will provide an
independent comparison to the results obtained by the FRANEC
code and by \citet{marigo02}. Since a clear
dependence of the mass-loss rate on the metallicity has still not
be identified, different mass-loss prescriptions should be tested
(Cristallo et al., in preparation). Finally, we note that the possible 
occurrence of the dual shell flash at the beginning of the AGB phase may 
also affect fluorine production and needs to be investigated in detail 
\citep[e.g,][]{campbell08}.

\section{Discussion and conclusions}\label{sect:maincnclsns}

We have shown that it is possible to reproduce the C and F 
abundances observed in
the CEMP-s star HE~1305+0132 via binary mass transfer from 
a companion by accretion
of 3-11$\%$ of the mass lost by the primary star during its AGB phase. The AGB
star should have dredged-up at least $\simeq$0.2\Msun~of its intershell material
into the convective envelope by means of the TDU. 
While rapidly rotating massive stars 
produce enough carbon and nitrogen to form CEMP stars, they do not appear 
to produce fluorine \citep{palacios05}. The binary formation 
scenario is thus favoured in the case of HE~1305+0132. In general, we 
predict that most CEMP 
stars formed by mass transfer from an AGB companion 
should also be FEMP stars, i.e., have [F/Fe] $>$ +1.
Hence, fluorine appears to be a useful discriminant between the 
different scenarios proposed for the formation of CEMP stars. 

During the preparation of this manuscript, another 
halo object highly enriched in fluorine was discovered, the 
planetary nebula BoBn~1 
\citep{otsuka08b}. The metallicity, as well as all the C and N abundances 
observed in 
this object are the same, within errors, as those of HE~1305+0132, which 
suggests that BoBn~1 has a close connection to CEMP stars, perhaps 
representing the evolutionary outcome of a CEMP star. On the other hand, the 
derived F abundance in this object is roughly a factor of three higher than that
obtained for HE~1305+0132. This observation can be explained via the binary 
scenario only if we consider 
the F yields we computed including the $^{13}$C-$^{14}$N pocket or 
the upper limit of the $^{18}$F($\alpha,p$)$^{21}$Ne reaction.
Following this indication, we multiplied the F yields by a 
factor of three in our stellar population model and calculated a probability of 
12\% for CEMP stars to have A($^{19}$F)$ \geq +4.75$. This result provides us with    
a possibility to alleviate the problem of the 
extremely small probability of the high F 
abundance assessed for HE~1305+0132.

\begin{acknowledgements} 
We thank an anonimous referee for her/his criticisms, which have much helped improving 
the clarity and focus of this paper. ML and RGI gratefully acknowledge the support of NWO. 
SC acknowledges the APAC national supercomputing facility. AIK acknowledges 
support from the Australian Research Council. 
RG acknowledges support by the Italian MIUR-PRIN06 Project
``Late phases of Stellar Evolution: Nucleosynthesis in Supernovae, AGB
stars, Planetary Nebulae''. TCB acknowledges partial support 
for this work from the National Science Foundation under grants AST 04-06784, 
AST 07-07776, and PHY 02-16783, Physics Frontier Center/Joint Institute for 
Nuclear Astrophysics (JINA). 
\end{acknowledgements}

%\bibliography{apj-jour,library}

\end{document}